\documentclass[twocolumn,showpacs,aps,prl,superscriptaddress,floatfix]{revtex4}
\usepackage{graphicx}
\usepackage{amsmath,amssymb,amsfonts,mathrsfs}
\usepackage{epic,eepic}
\usepackage{graphics}
\usepackage{epsfig}
\usepackage{epsfig}
\usepackage{subfigure}
\usepackage{epic,eepic}
\usepackage{graphics}
\usepackage{epsfig}
\usepackage{subfigure}
\usepackage[latin1]{inputenc}
\usepackage{pstricks,pst-node,pst-text,pst-3d}
\usepackage{amsmath}
\usepackage{graphicx}
\usepackage{pst-grad}
\usepackage{pst-vue3d}
\usepackage{graphicx}
\usepackage{pst-grad}
\usepackage{pst-vue3d}
\usepackage{verbatim}
\usepackage{amsmath}
\usepackage{amsfonts}
\usepackage{amssymb}
\usepackage{bm}
\usepackage{color}
\usepackage{graphicx}
\usepackage{verbatim}
\usepackage{subfigure}
\usepackage{amsmath}
\usepackage{bm}
\usepackage{amssymb}
\usepackage{natbib}
\usepackage{color}
\usepackage{ulem}
\usepackage{soul}

\newpsstyle{GradGrayWhite}{fillstyle=gradient,    gradbegin=blue,gradend=white,linewidth=0.1mm}
\newpsstyle{GradGrayWhite}{fillstyle=gradient,    gradbegin=blue,gradend=white,linewidth=0.1mm}

\newcommand{\be}{\begin{equation}}
\newcommand{\ee}{\end{equation}}

\input{tcilatex}

\begin{document}

\title{Stick-Slip Sliding of Water Drops on Chemically Heterogeneous Surfaces}
\author{S. Varagnolo$^{1}$, D. Ferraro$^{1}$, P. Fantinel$^{1}$, M. Pierno$^{1}$, G. Mistura$^{1}$, G. Amati$^{2}$, L. Biferale$^{3}$, M. Sbragaglia$^{3}$\\
$^{1}$ CNISM and Department of Physics and Astronomy ``G. Galilei'', University of Padova, Via Marzolo 8, 35131 Padova, Italy\\
$^{2}$ Department of Physics, University of ``Tor Vergata'', Via della Ricerca Scientifica 1, 00133 Rome, Italy\\
$^{3}$ Department of Physics and INFN, University of ``Tor Vergata'', Via della Ricerca Scientifica 1, 00133 Rome, Italy}

\begin{abstract}
We present a comprehensive study of water drops sliding down chemically heterogeneous surfaces formed by a periodic pattern of alternating hydrophobic and hydrophilic stripes. Drops are found to undergo a stick-slip motion whose average speed is an order of magnitude smaller than that measured on a homogeneous surface having the same static contact angle. This motion is the result of the periodic deformations of the drop interface when crossing the stripes. Numerical simulations confirm this view and are used to elucidate the principles underlying the experimental observations.
\end{abstract}

\maketitle

Controlling and predicting the equilibrium and dynamical properties of drops on chemically heterogeneous surfaces is a major scientific challenge, relevant not only for fundamental research \cite{Bonn09,Podgorskietal01,Hodges04,LimatStone05} but also for an ample variety of applications, particularly in droplet based microfluidics, a rapidly growing interdisciplinary field of research combining soft matter physics, biochemistry and microsystems engineering \cite{Applications2}. Understanding the liquid dynamics is essential to design new smart coatings which, for example, guide the wetting drops along certain directions \cite{Gauetal99,Ledesmaetal11}. When drops are deposited on a surface functionalized with stripes of alternating wettability, they may assume elongated shapes, which are characterized by different contact angles in the directions perpendicular and parallel to the stripes. This morphological anisotropy has been the object of intense scrutiny in a variety of situations \cite{Pompe00,Buehrle02,Moritaetal05,Semprebonetal09,Jansenetal12a}. Studies about the sliding of drops on striped surfaces also report an anisotropic dynamics: drops move faster along the alternating stripes than across them \cite{Moritaetal05,Suzukietal08} and velocity fluctuations take place along this latter direction \cite{Suzukietal08}. Numerical simulations \cite{KusumaatmajaYeomans07,Wangetal08,Herdeetal12} and theoretical studies \cite{ThieleKnobloch06b,ThieleKnobloch06a,Beltrame09} also point to the occurrence of a stick-slip motion when the drop crosses stripes of different wettability. The observed anisotropies are usually explained in terms of the energy barrier the drop must overcome in the direction perpendicular to the stripes, while it is ideally free to move in the direction parallel to them.

Hereafter, we present the first experimental evidence of a stick-slip motion of water drops sliding down heterogeneous hydrophilic/hydrophobic surfaces having parallel stripes with a large wettability contrast of about 70\textdegree. We find that the mean speed of the sliding drop is strongly affected by the patterning details, with a slowing down that can reach up to a factor $10$ with respect to the corresponding homogeneous coating with the same static contact angle. We also show that such a big change in the dynamical evolution is due to a different balancing between capillary, viscous and body forces in the presence of stick-slip.

\begin{figure}[tbp]
\includegraphics[scale=0.42]{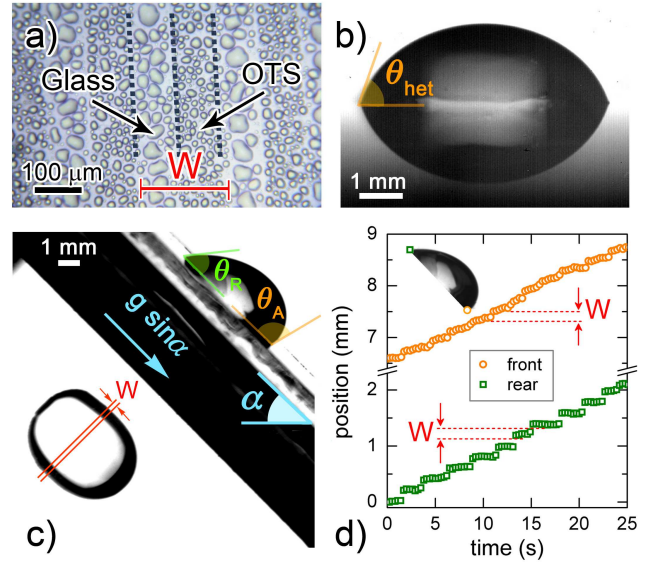}
\caption{(Color online) a) Condensation figure on a heterogeneous surface: big (small) water drops form on hydrophilic glass (hydrophobic OTS). Parallel stripes with periodicity $W$ can be easily distinguished according to the size and distribution of the drops; b) Static profile of a water drop on the heterogeneous surface; c) Lateral view (top right) and contact line view (bottom left) of the sliding drops on inclined substrates; d) Typical Stick-Slip time law for the front and rear contact points of the sliding drop.}
\label{fig:1}
\end{figure}
Chemically heterogeneous surfaces are produced through micro contact printing starting from the realization of masters with rectangular grooves by photolithography. Such masters are replicated in polydimethilsiloxane (PDMS) to obtain the stamp \cite{Tothetal11} for the printing of a solution of octadecyltrichlorosilane (OTS) in toluene on a glass substrate. The result is a surface presenting alternating hydrophobic (OTS regions) and hydrophilic stripes (uncoated glass regions). Sample characterization is performed through atomic force microscopy and the condensation figure method as shown in Fig.~\ref{fig:1}a. Parallel stripes of equal width and different wettability can be clearly evinced having a surface roughness of less than 10 nm measured over an area of 10 $\mu \text{m}$ $\times$ 10 $\mu \text{m}$. The resulting pattern has a periodicity $W\sim 200\,\mu \text{m}$. The printed pattern is also analyzed in terms of contact angle measurements through the Cassie-Baxter equation \cite{Bonn09}. Figure~\ref{fig:1}b shows the profile of a drop deposited on the heterogeneous surface. Due to the size of the stripes and the drop volume, the effects of morphological anisotropies \cite{Moritaetal05, Jansenetal12a} are not very pronounced, yielding an equilibrium contact angle parallel to the stripes $\theta _{\textrm{het},\|}= 72^{\circ} \pm 3^{\circ}$, which fits well with the Cassie prediction $\theta_{\textrm{het}}= 75 ^{\circ} \pm 2^{\circ}$ calculated from the pattern details.

Water drops of the desired volume $V$ ($\approx $ 30 $\mu\textrm{l}$) are gently deposited on the sample surface through a vertical syringe pump. The sample is placed on a tiltable support, whose inclination can be set with a 0.1\textdegree\ accuracy. A mirror mounted under the sample stage at 45\textdegree\ with respect to the surface allows to simultaneously view the contact line and the lateral side of the drop \cite{LeGrandDaerr05}, as shown in Fig.~\ref{fig:1}c. The drop is lightened by two LED back lights: one of them is placed in front of the camera and the other is fixed above the sample stage, to illuminate the drop from the top. Experimentally, we first fix the inclination of the plane and then deposit the drop. Acquired images, where drops appear dark on a light background, are analyzed through a custom made program \cite{Ferraroetal12}.\\
\begin{figure}[tbp]
\includegraphics[scale=0.52]{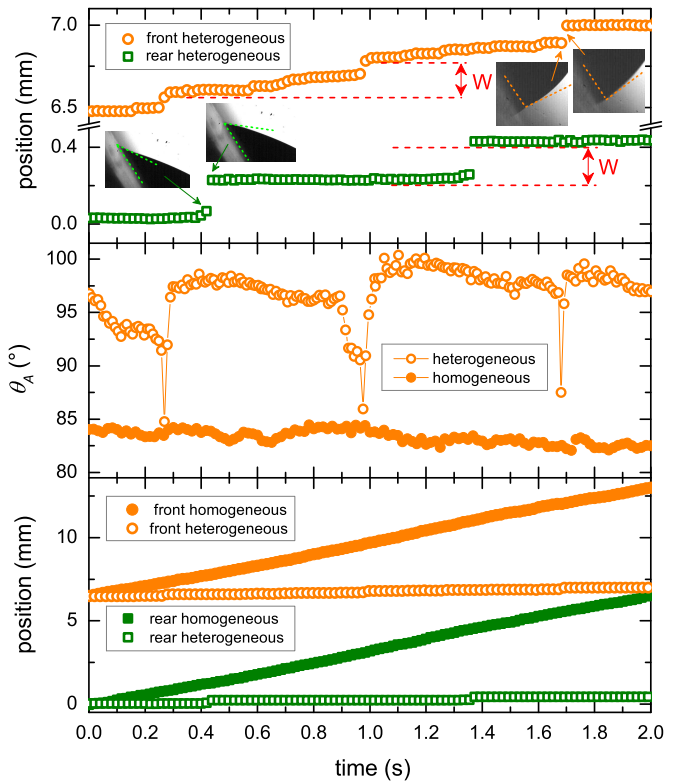}
\caption{(Color online). Top: time evolution of the position of the rear and the front contact points of a drop sliding down the heterogeneous surface tilted by $\protect\alpha $=56\textdegree. Snapshots show the different contact angles before and after the jumps. Middle: variation of the advancing contact angle $\protect\theta _{A}$ of the same drop and of a different one sliding down an homogeneous surface with a $\protect\theta_{\textrm{eq}}$=71\textdegree\ inclined by the same $\alpha$. Bottom: corresponding evolution of the frontal and rear contact points on heterogeneous and homogeneous surfaces.}
\label{fig:sl_exp}
\end{figure}
Fig.~\ref{fig:1}d shows the positions of the front and rear contact points of a drop sliding down the heterogeneous surface perpendicularly to the stripes. This is more clearly seen in the top panel of Fig.~\ref{fig:sl_exp}, which shows the drop dynamics in a time window expanded by a factor $\approx $10 with respect to the time scale of Fig.~\ref{fig:1}d. The drop clearly advances with a stick-slip behavior, with jumps of the order of the pattern periodicity $W$. At the beginning of the time frame considered, the rear contact line is pinned while the front line slowly advances until around $t\approx $ 0.25 s when it suddenly jumps forward. In correspondence, a drop in $\theta _{A}$ is observed (see the middle panel of Fig.~\ref{fig:sl_exp}). Subsequently, the front contact line hardly moves, while the rear line at $t\approx $ 0.4\ \textrm{s} jumps by a distance equal to $W$. The process then repeats itself. This peculiar dynamics has never been observed experimentally. In previous studies this stick-slip motion has not been seen probably because the sliding occurred across stripes having a small wettability contrast \cite{Suzukietal08}. A stick-slip has also been reported in the spreading of a drop of increasing volume along the direction orthogonal to the stripes while in the other direction the contact line moves at a constant speed \cite{LeopoldesBucknall05}.

To better understand this motion, we have repeated the measurement with a glass surface coated with a homogeneous OTS layer deposited from the vapor phase which presents a static contact angle $\theta _{\textrm{eq}}\approx 71^{\circ} \pm 2^{\circ}$, very close to the Cassie angle of the heterogeneous surface. Furthermore, the inclination angle of the OTS surface and the water drop volume are the same as in the case of the heterogeneous surface. Statically, then, the two systems look pretty much the same, but dynamically their behavior is quite different. On the homogeneous
substrate, as expected, the motion is uniform as shown in the bottom panel of Fig.~\ref{fig:sl_exp} and $\theta_{A}$ practically does not change (see middle panel). Moreover, the drop velocity on the homogeneous surface is about 10 times larger than the average velocity on the heterogeneous surface (see bottom panel).

\begin{figure}[tbp]
\includegraphics[scale=0.41]{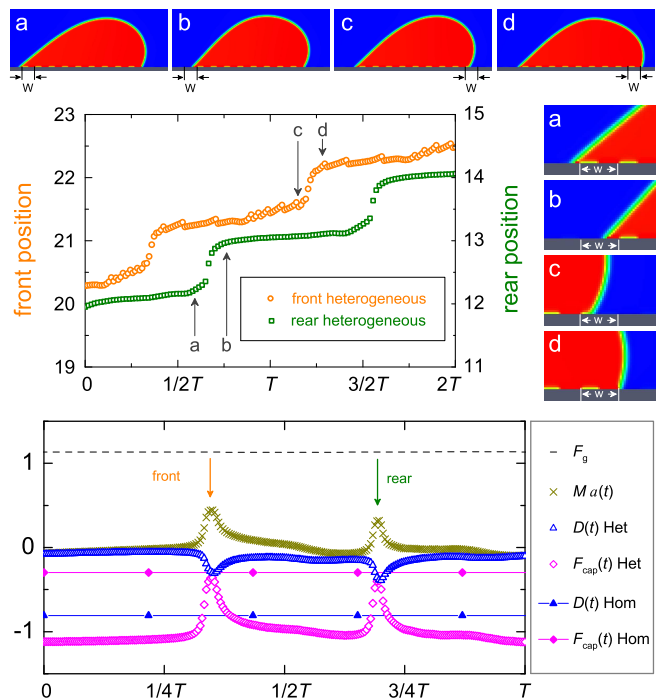}
\caption{(Color online) Top: time evolution in dimensionless units (see text for details) of the front (left axis) and rear (right axis) contact points of a 2D drop on the heterogeneous substrate. The horizontal sequence (on the top) shows the whole drop in terms of density snapshots: red (blue) color is associated to high (low) density regions. The vertical sequence (on the right) reports the corresponding enlargements to better appreciate the position of the contact line. Consistently, the yellow (red) areas at the drop base represent hydrophobic (hydrophilic) regions.
Bottom: time evolution of the various terms in the balance
equation (\protect\ref{eq:balance}) for the heterogeneous and the
homogeneous case chosen to display the same static morphology.}
\label{fig:num}
\end{figure}

This physical picture is corroborated with the use of numerical simulations based on the lattice Boltzmann models \cite{Moradi11,Sbragaglia06,Hyvaluomaetal07}, which allow to simulate the continuity and the Navier-Stokes equations with capillary forces at the interfaces. A detailed description of the model can be found in other publications \cite{CHEM09,SC1,Premnath}. We shall investigate a two-dimensional situation: this is the optimal choice to retain the essential features of the stick-slip dynamics \cite{KusumaatmajaYeomans07,Herdeetal12,Beltrame09} and, at the same time, achieve a reasonably large resolution to resolve the hydrodynamic equations inside the drop. At variance with respect to other numerical simulations done in the quasistatic limit \cite{KusumaatmajaYeomans07,Herdeetal12} or  referring to single contact line motion \cite{Wangetal08}, we will show explicitly how the different terms in the equations of motion balance in the stick-slip dynamics. The output of the numerical simulations (i.e. density, velocity, forces) is integrated over the drop volume and made dimensionless with respect to the time-length scales of the problem. We reproduce the following force-balance equation
\begin{equation}
M a(t)=D(t)+F_{g} +F_{\textrm{cap}}(t)  \label{eq:balance}
\end{equation}
where $a(t)$ is the acceleration of the drop of mass $M$ and $F_{g}$ is the gravitational force. The term $F_{\textrm{cap}}$ accounts for the internal pressure forces due to the curvature distortion of the interface ($F_{\textrm{cap}}=0$ for a spherical interface) while the function $D(t)$ quantifies the dissipation inside the drop. The simulation parameters are chosen in such a way to reproduce the experimental conditions investigated in Fig.~\ref{fig:sl_exp}. The observed stick-slip motion is captured in the numerical simulations as shown in the top graph of Fig.~\ref{fig:num}, which displays the time evolution of the positions of the front and rear contact points normalized to $W$, and $T$ indicates the time period required for a displacement equal to $W$. From the density snapshots it is possible to determine precisely the positions where the contact line pins/depins with a resolution that is impossible to achieve in an experiment. The rear contact line gets pinned (snapshots a) before entering the hydrophobic area (yellow region at the base of the drop). A sudden jump makes it overcome the local energy barrier and enter the hydrophilic area (red region at the base of the drop)  (snapshots b). The front contact line, in turn, stops before entering the hydrophobic regions. As the drop pins with an increasing advancing angle, it slowly penetrates through the hydrophobic area (snapshots c), similarly to what happens in Fig.~\ref{fig:sl_exp}. A sudden jump follows when the front enters the hydrophilic area (snapshots d). We have also extended the simulations to the heterogeneous surface investigated in \cite{Suzukietal08}, which presents a lower wettability contrast of 10\textdegree\, and found only periodic oscillations in the velocity without the occurrence of a stick phase in agreement with the experiment.\\The bottom graph of Fig.~\ref{fig:num} presents the analysis of the balance equation (\ref{eq:balance}) for a time frame of a period $T$ of the stick-slip dynamics and compares it to the homogeneous case, for the same $Bo$. The gravitational force $F_g$ is constant in time. When the drop is pinned, $F_g$ is almost balanced by $F_{\textrm{cap}}$ (snapshots c). Immediately after, the front contact line jumps forward and the drop depins ($F_{\textrm{cap}} \rightarrow 0$): the drop experiences a localized acceleration $a(t)$ with a consistent dip in the dissipation function. The drop gets pinned again until the rear contact line jumps forward. The homogeneous case is instead stationary: the energy provided by $F_g$ is almost entirely transferred into dissipation, apart from the deformation of the interface which causes a term $F_{\textrm{cap}}$ smaller by a factor $\approx $10 with respect to the heterogeneous case. Overall, we see that the effective dissipation in the heterogeneous case is strongly suppressed as compared with the stationary homogeneous case.  This is because the large wettability contrast causes additional energy to be stored in the non-equilibrium configuration of the drop which can pin before the contact lines jump forward.

\begin{figure}[tbp]
\includegraphics[scale=0.57]{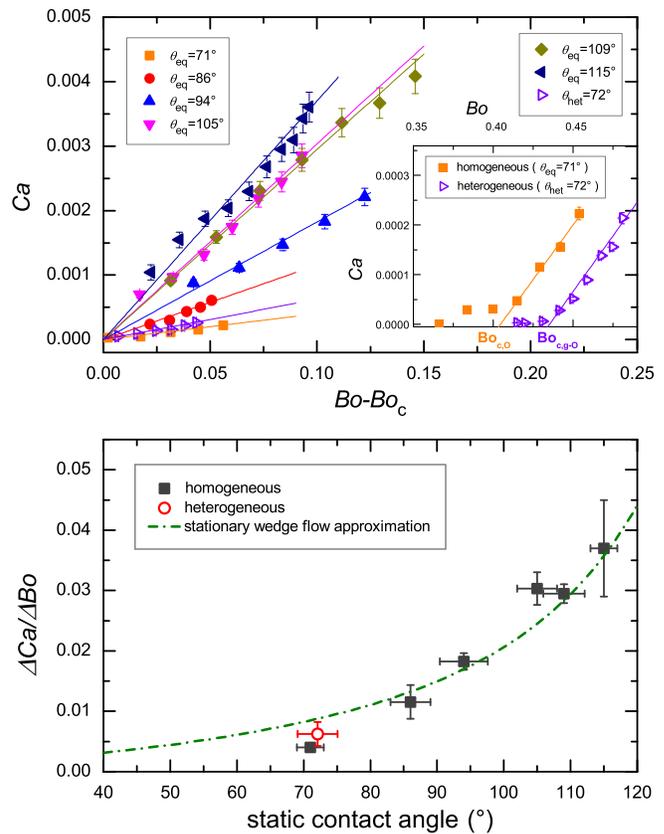}
\caption{(Color online) Top: relation between the Capillary number ($Ca$) and the Bond number ($Bo$) for both homogeneous (filled symbols) and heterogeneous (open symbols) surfaces. All the Bond numbers have been rescaled by subtracting the critical Bond number $Bo_{c}$, which is larger for the heterogeneous case (see inset). Bottom: the slope obtained from the linear fits of the $Ca$ vs. $Bo$ curve is reported as a function of the static contact angle. The dashed line is the scaling law predicted by (\ref{eq:scaling}), calculated for small drops sliding down homogeneous surfaces with a wedge dissipation as the dominant dissipative contribution \cite{HuHScriven71,Kimetal02}.}
\label{fig:Ca_Bo}
\end{figure}

Our next goal is to quantify the dependence of the drop average speed on the inclination angle. Again, this will be achieved by comparing the homogeneous and heterogeneous cases. We have performed measurements with different coatings of glass slides produced with different molecules and methods, obtaining contact angles ranging from $\theta _{\textrm{eq}}=71^{\circ} \pm 2^{\circ}$ to $\theta _{\textrm{eq}}=115^{\circ} \pm 2^{\circ}$. The sliding of water drops of volume $V$ down these surfaces has been studied at different inclinations $\alpha$. For angles above the sliding angle, the drop moves at the constant speed $U$, which increases with the inclination angle. Such motion is the result of a balance between the in-plane component of the drop weight and the viscous resistance, plus a capillary force related to the nonuniformity of the contact angle along the drop perimeter \cite{Podgorskietal01,LeGrandDaerr05}. For small drops, this leads to the following scaling law between the capillary number $Ca=\eta U/\gamma$, where $\gamma$ is the surface tension and $\eta$  the viscosity, and the Bond number $Bo=(3V/4\pi )^{2/3}\rho g\sin\alpha /\gamma$, $\rho$ being the density \cite{Kimetal02}:
\begin{equation}
Ca \sim (Bo-Bo_{c})/c(\theta )  \label{eq:scaling}
\end{equation}
where the function $c(\theta )$ is related to the solution of the hydrodynamic flow \cite{HuHScriven71} in the wedge. In principle, the angle $\theta $ is the dynamical contact angle. Nevertheless, it is plausible to approximate $\theta \approx \theta _{\textrm{eq}}$ either when dynamic contact angles do not deviate severely from the equilibrium contact angle or when the mean of the advancing and receding contact angles is close to $\theta _{\textrm{eq}}$ \cite{Kimetal02}. We therefore get $c(\theta)=(1-\cos^{2}\theta _{\textrm{eq}})(\theta_{\textrm{eq}}-\sin \theta _{\textrm{eq}}\cos \theta _{\textrm{eq}})$ multiplied by an angular function relating the radius of the bottom contact area to the volume \cite{Kimetal02}.
Data displayed in the inset of Fig.~\ref{fig:Ca_Bo} refer to drops of $V$ $\approx $ $30$ $\mu \textrm{l}$ sliding on both the heterogeneous and the homogeneous surfaces with the same static contact angle. The $Bo_{c}$ value is deduced by extrapolating the linear fit of the sliding points to $Ca=0$. It is evident that the two surfaces display a different dynamic behavior, the heterogeneous one being characterized by a larger pinning. To better compare the sliding on the various surfaces, in the top panel of Fig.~\ref{fig:Ca_Bo} we plot all the data in terms of $(Bo-Bo_{c})$. In this way, the points lie on straight lines passing through the origin, whose slopes depend on the surface wettability. The slopes as a function of $\theta _{eq}$ are displayed in the bottom panel of Fig.~\ref{fig:Ca_Bo}: the agreement between the angular dependent prefactor of equation (\ref{eq:scaling}) and the experimental data is quite reasonable for all the investigated surfaces, including the heterogeneous one. This indicates that the effects of the heterogeneous patterning can be readsorbed in a renormalized value of the $Bo_c$, representing the increase of the static energetic barrier that must be overwhelmed by gravity before the drop starts to move, an observation that bears similarities with the results discussed in \cite{Herdeetal12}.\\
In summary, we have observed both experimentally and numerically that a water drop, sliding across alternating stripes having a large wettability contrast, undergoes a characteristic stick-slip motion. The average speed of this nonlinear motion is an order of magnitude smaller than that measured on a homogeneous surface having the same static contact angle. The slow down is the result of the pinning-depinning transition of the contact line which causes energy dissipation to be localized in time and large part of the driving energy to be stored in the periodic deformations of the contact lines when crossing the stripes. This suggests that the motion of drops can be passively controlled by a suitable tailoring of the chemical pattern.
Experiments are under way to see how the shape of the tiles forming the periodic, heterogeneous patterns affects the drop dynamics.\\

We are particularly grateful to M. Brinkmann and C. Semprebon for useful
discussions. MS, LB, GA and MP kindly acknowledge funding from the European
Research Council under the Europeans Community's Seventh Framework Programme
(FP7/2007-2013) / ERC Grant Agreement no[279004].

\end{document}